\documentclass{aa}
\usepackage{graphics,times,psfig}

\begin{document}

\thesaurus{03.13.4, 11(09.2,12.1)}

\title{Sagittarius, a dwarf spheroidal galaxy without dark matter?}

\author{M.A. G\'omez-Flechoso, R. Fux and L. Martinet}

\institute{Geneva Observatory, Ch. des Maillettes 51, CH-1290 Sauverny, 
Switzerland}

\authorrunning{G\'omez-Flechoso et al.}
\titlerunning{Sagittarius}
\date{Received / Accepted}
\maketitle

%%%%%%%%%%%%%%%%%%%%%%%%%%%%%%%%%%%%%%%%%%%%%%%%%%%%%%%%%%%%%%%%%%%%%%%%%%%%%%
\begin{abstract}
The existence of dwarf spheroidal galaxies with high internal velocity 
dispersions orbiting in the Milky Way raises questions about their
dark matter content and lifetime. In this paper, we present an 
alternative solution to the dark matter dominated satellites
proposed by Ibata \& Lewis (\cite{Ibata98}) for the Sagittarius
dwarf galaxy. 
 We performed simulations of two kinds of N-body satellites: 
the first models (f-models) could correspond to satellites with high dark matter
content and they represent initially isolated models. The second models (s-models) 
have either low or negligible dark matter content and they are constructed in
a tidal field.
In spite of being on the same orbits, the s-models are able to produce a 
better agreement with some observational constraints concerning Sagittarius.
From our simulations, we can also infer that Sagittarius is in the 
process of being disrupted.
%%%%%%%%%%%%%%%%%%%%%%%%%%%%%%%%%%%%%%%%%%%%%%%%%%%%%%%%%%%%%%%%%%%%%%%%%%%%%%

\end{abstract}
%%%%%%%%%%%%%%%%%%%%%%%%%%%%%%%%%%%%%%%%%%%%%%%%%%%%%%%%%%%%%%%%%%%%%%%%%%%%%%

\section{Introduction}

The current interest for dwarf spheroidal galaxies (hereafter DSphs) 
as satellites 
of the Milky Way is raised by two fundamental questions concerning
their evolution: the real content of dark matter (DM) in these objects 
(see, for example, Mateo \cite{Mateo94},\cite{Mateo97},\cite{Mateo98b};
Piatek \& Pryor \cite{Piatek95}; 
Burkert \cite{Burkert97} and references therein)
and their implication in the hierarchical formation of the galactic 
halo (e.g., Johnston et al. \cite{Johnston96}).

The Sagittarius (Sgr) dwarf galaxy is the closest known satellite galaxy to
the center of the Milky Way, $R_{GC} \sim 16$ kpc (Ibata et al. \cite{Ibata95},
\cite{Ibata97}). Due to its proximity we can expect from its
study an additional contribution to our understanding of DSphs in general.
Since the announcement of its discovery by Ibata et al. (\cite{Ibata94})
the structure and evolution of Sgr have been extensively 
discussed and simulated by various authors: Ibata et al. (\cite{Ibata95}), 
Johnston et al. (\cite{Johnston95}), Vel\'azquez \& White (\cite{Velazquez95}), 
Whitelock et al. (\cite{Whitelock96}),
Mateo et al. (\cite{Mateo96}), Alard (\cite{Alard96}),
Ibata et al. (\cite{Ibata97}), 
Edelsohn \& Elmegreen (\cite{Edelsohn97}), Layden
\& Saragedini (\cite{Layden97}), Ibata \& Lewis
(\cite{Ibata98}), Mateo et al. (\cite{Mateo98}). 
Important problems concerning this system are: 1) its possible lifetime
before dissolution, 2) the possible presence of DM in it. A
critical point in the studies mentioned above is the question whether Sgr is in
virial equilibrium or not. 

\begin{table}
\caption{Observational parameters of Sagittarius dwarf galaxy (Ibata et 
al. \cite{Ibata97})}
\label{Sgr_tbl}
\begin{minipage}{5cm}
\begin{tabular}{ll}
\hline
Parameter& \\
\hline
$r_{hb}$ & 0.55 kpc\\
$\sigma_o$ & 11.4 km/s \\
$\mu_{oV}$ & 25.4 mag/arcsec$^2$ \\
$L_{t}$ & $\geq 10^7$ L$_{\odot}$ \\
$(M/L)_o$ & 50 M$_{\odot}/$L$_{\odot}$ \\
$M_{t}$ & $> 10^9$ M$_{\odot}$ \\
$(l,b)$\footnote{Galactic coordinates} & (5.6$^o$, -14$^o$) \\
$(U,V,W)$\footnote{Galactic velocities} & (232,0:,194)$\pm$ 60 km/s \\
$d_{\odot}$\footnote{Heliocentric distance} & 25 kpc \\
$R_{GC}$\footnote{Galactocentric distance} & 16 kpc \\
$v_r$\footnote{Radial velocity} & 171 km s$^{-1}$\\
$(dv/db)$\footnote{Gradient of the radial velocity} & -3 km s$^{-1}$/degree \\
\end{tabular}
\end{minipage}
\end{table}

 Usually, the total inferred mass of the DSph is calculated by assuming 
it is in virial equilibrium\footnote{In this paper, the {\it inferred mass} is 
always obtained by assuming a virial equilibrium for the satellite.}. 
In this case, 
the central mass-to-light ratio depends on the velocity
dispersion through the equation (Richstone \& Tremaine \cite{Richstone86})

\begin{equation}
\left(\frac{M}{L}\right)_o= \eta \frac{9 \sigma^2_o}{2 \pi G \mu_o r_{hb}}
\label{ML}
\end{equation}

\noindent
where $\eta$ is near unity for a wide variety of models, $\sigma_o$ the
central velocity dispersion, $\mu_o$ the central surface brightness and
$r_{hb}$ the half-brightness radius. 
The analysis of the validity of Eq. (\ref{ML}) for evolving DSphs
has been studied in detail in the present context 
by Kroupa (\cite{Kroupa97}, hereafter K97).

In Table \ref{Sgr_tbl}, we have summarized the parameters
of Sgr DSph measured by Ibata et al. (\cite{Ibata97}). For the values
of $\sigma_o$, $r_{hb}$ and $\mu_{oV}$ given in Table \ref{Sgr_tbl}, 
those authors obtain a high central
mass-to-luminosity ratio $(M/L)_o= 50$ M$_{\odot}/$L$_{\odot}$ by using
the Eq. (\ref{ML}). Therefore, the total inferred mass
assuming virial equilibrium, $M_{t}$, is DM dominated. The values
of total luminosity, $L_{t}$, and the total inferred mass, $M_{t}$, are
also listed in Table \ref{Sgr_tbl}.
However, the values of $(M/L)_o$ and $L_t$ suggested 
by Mateo et al. 
(\cite{Mateo98}) are different. These authors have discovered a tidal
extension of Sgr, which implies $L_t \leq 5.8 \times 10^7$ L$_{\odot}$ and,
as a consequence, the $(M/L)_o$ ratio could 
decrease to 10 M$_{\odot}/$L$_{\odot}$.
Adopting the structural 
and orbital characteristics usually assumed (given in Table \ref{Sgr_tbl}), 
Ibata et al. (\cite{Ibata97}) 
recently suggested that such a satellite galaxy is expected to
be tidally disrupted and destroyed after several pericentric passages, 
unless a significant quantity of DM is present inside it, as
inferred assuming virial equilibrium. 
The observations
of the Sgr globular clusters give an age spread of its constituents $> 4$ Gyrs
and the youngest globular cluster (Terzan 7) is 9-13 Gyrs old (Montegriffo
et al. \cite{Montegriffo98}). The fast disruption obtained by the 
numerical methods raise the question about the age and the
dynamical history of Sgr. However, it must be emphasized that the initial 
time $t=0$ 
of the simulations does not necessarily coincide with the time of
formation of the oldest constituents of the satellite.

The partial formation of the
galactic halo by hierarchical processes (accretion of small galaxies) has 
recently received an increase of interest, stimulated 
for instance by the 
investigation by Lynden-Bell \& Lynden-Bell (1995) on the reality 
of streams (Lynden-Bell \cite{Lynden82})
in the close environment of the Milky Way, among them the
well known Magellanic Stream. 
DSphs seem to belong to one or another of these streams and their 
evolution clearly depends on their environment. 
This scenario of satellite formation involves a low DM content
for the DSphs (Barnes \& Hernquist \cite{Barnes92}; Kroupa \cite{Kroupa98b}). 
However,
the only simulations of no-dark matter satellite galaxies 
able to survive in a tidal field are those of K97 
and Klessen \& Kroupa (\cite{Klessen98}, hereafter KK98). These authors have studied a 
region in the parameter space ($M_{sat}, r_{sat}$), where
$M_{sat}$ and $r_{sat}$ are the mass and a typical radius of the 
satellite, 
and they have obtained a 
residual satellite from a more massive one.
Nevertheless, their remnant satellite 
galaxies are fainter than a typical DSph, unless the true $(M/L)_{\mbox{real}} < 3$. 

These circumstances make questionable the
maintenance of dynamical equilibrium and consequently the virial estimation
of mass for these systems.

In this paper we present self-consistent N-body simulations of 
Sgr and a more accurate and plausible scenario of its evolution, which 
suggests that Sgr is likely to be able to survive for a long time 
($6-10$ Gyr)
by orbiting in the Galaxy without being dominated by DM.

In Sect. 2, we present the model of the Milky Way used in our simulations.
In Sect. 3, we describe the different models of the satellite galaxy and
corresponding scenarios of interaction with the Galaxy. 
In Sect. 4 some
numerical considerations are given. In Sect. 5, we present the results of our
simulations for the different chosen 
scenarios. These results are discussed in 
Sect. 6 in connection with recent conjectures by Kroupa (\cite{Kroupa98}, 
hereafter K98) on
the parameter space possible for progenitors of 
surviving DSphs without DM. Finally, a summary is given in Sect. 7.

\section{The model of the Milky Way}
\label{galaxy}

\begin{figure}[t]
\centerline{\psfig{figure=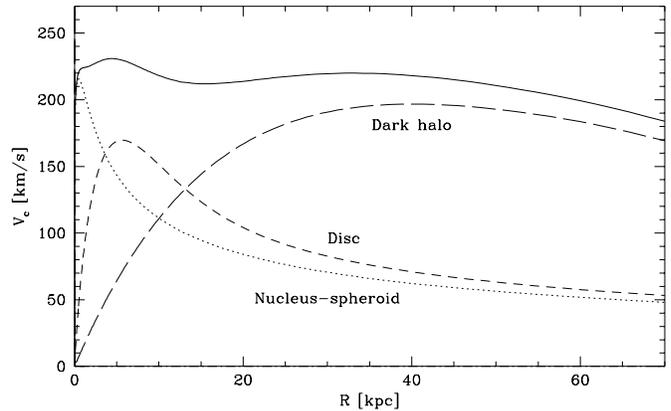,width=8.8cm}}
\caption[]{Rotation curve of the initial Milky Way model (full line), with the
           contributions of each component}
\label{f1}
\end{figure}

The initial model of the primary galaxy, representing the Milky~Way, is based
on the axisymmetric initial conditions adopted by Fux (\cite{Fux97}) 
in his N-body
modeling of the Galaxy. The mass distribution is divided in three
components: (i) an oblate stellar nucleus-spheroid with a spatial density
$\rho\propto r^{-1.8}$ in the central region, in agreement with the
near-infrared observations of the inner bulge (Becklin \& Neugebauer 
\cite{Becklin68};
Matsumoto et al. \cite{Matsumoto82}), 
and $\propto r^{-3.3}$ outside the bulge, as the number
density counts of stellar halo objects (e.g. Preston et al. \cite{Preston91} for
RR Lyrae; Zinn \cite{Zinn85} for globular clusters), 
(ii) a double exponential stellar
disc, with radial and vertical scale lengths $h_R=2.5$~kpc and $h_z=250$~pc
and (iii) an oblate dark halo with an exponential profile ensuring a roughly
flat rotation curve out to $R=40$~kpc. Both oblate components have a
flattening $c/a=0.5$. The total mass of the luminous component is
$M_L=8.25 \times 10^{10}$ M$_{\odot}$. 
The model is identical to Fux's model m04t0000, except
that the dark halo is more extended, with a scale length $b=13$~kpc and a
total (untruncated) mass $M_{\rm DH}=2.4\times 10^{11}$~M$_{\odot}$, and that
the truncation radius is moved further away at $R_{\rm c}=70$~kpc. The chosen
truncation radius ensures
a non-vanishing density everywhere along the orbit of the satellite galaxy. 
 The resulting rotation curve of the model is shown in Fig.~\ref{f1}.

\par The initial kinematics is obtained by solving the hydrodynamical
Jeans equations, assuming an isotropic velocity dispersion for the disc
component and a anisotropic but centrally oriented velocity ellipsoid for the
other components. The velocity distribution of the nucleus-spheroid and
dark halo is generated from a 3D generalization of the Beta distribution
which limits the number of escaped particles and thus greatly improves the
equilibrium of the isolated dynamical model. More details are given in
Fux~(\cite{Fux97}).

\section{Satellite models}

 The effects that a DSph suffers when it is accreted 
on a primary galaxy strongly depend on its structure and 
its orbit. 

We have considered two kinds of satellite models. They correspond to two
scenarios of interaction between the Galaxy and the satellite.

 In the first models (f-models), 
we represent the satellite as a {\it standard} King's
sphere (King \cite{King66}), which matches the observational 
constraints for the total
virial mass, $M_t$, the core radius, $r_o$, and the central velocity 
dispersion, $\sigma_o$, of the Sgr DSph. 
In this case,
the effects on the DSph are maximal because the satellite, originally 
in an isolated situation, suddenly undergoes strong tidal perturbations.

 Our alternative scenario (s-models) assumes that
either the DSph is formed in
the tidal tail of another major accretion event and therefore it is built 
in equilibrium with the environment, or
the DSph falls slowly from a quasi-isolated situation to a tidal 
field region and it has enough time to readjust itself to the environmental
forces. In both situations, we begin our simulations when the 
satellite has already reached the equilibrium with the dense environment and,
therefore, 
the life time of the DSph orbiting in the galaxy is expected  to be longer 
than in our first scenario.
In the present case, the satellite is modelled by a {\it modified} King's 
sphere (G\'omez-Flechoso \& Dom\'{\i}nguez-Tenreiro \cite{Gomez97}), which takes
into account the tidal potential produced by the primary galaxy. 

 We have checked several orbits in order to compare the observational
features of the Sagittarius dwarf 
and the numerical results in both fast and slow accreting scenarios. The 
orbits we have chosen reproduce the present position, $(l,b)$, and 
galactocentric velocity,
$(U,V,W)$, of 
Sagittarius DSph in the observational range (Table \ref{Sgr_tbl}), 
and, therefore, the simulated orbits are polar, as suggested by the 
void component 
of the proper motion parallel to the Galactic Plane.
The orbits are either low
eccentricity orbits in the central part of the Galaxy or high eccentricity 
orbits.

 The central mass-to luminosity ratio of the models has been calculated 
by using the 
Eq. (\ref{ML}). The central velocity dispersion $\sigma_o$ involved in this
equation has been measured along the line-of-sight. We have removed 
the particles with the largest radial velocities (relative to
the radial velocity of the center of mass of the satellite) to prevent 
contamination by outlying particles. We have only considered
those particles with projected distance to the center of the satellite 
smaller than 0.5 kpc.

\subsection{f-models}

 The distribution function of an isolated galaxy fulfills the collisionless
Boltzmann equation. We can represent an isolated DSph as a solution of this 
equation. King's spheres are an example of that.

 If an initially isolated DSph reaches the inner regions of 
a galaxy within a short timescale, 
it has no time 
to modify its internal structure.
In this case the satellite maintains its isolated King's 
sphere distribution function at the beginning of the simulation.
Once in the inner orbit, the satellite will
evolve quickly, because tidal forces are 
strong in these regions. 
This scenario is unrealistic, because,
in reality, a satellite which reaches the denser parts of the galaxy 
has suffered 
the influence of the galaxy potential for a long period of time.
However, we have run 
simulations in such a case because they correspond to 
the common initial conditions assumed in the
literature (Johnston et al. \cite{Johnston95}, 
Oh et al. \cite{Oh95}, 
Vel\'azquez \& White \cite{Velazquez95}, 
Johnston et al. \cite{Johnston96}, 
Edelsohn \& Elmegreen \cite{Edelsohn97}, 
K97, KK98, 
Ibata \& Lewis \cite{Ibata98}) and in order to compare the results 
with those of the more realistic s-model 
satellites.

 The parameters of the King's model we have selected for the satellite model 
in a fast accreting scenario are given in Table \ref{tbl1} and they have
been chosen to reproduce the present characteristics of Sagittarius, as 
was done in other simulations (Vel\'azquez \& White \cite{Velazquez95}).
The total mass of the satellite model corresponds to the virial mass 
inferred for Sgr. Therefore, if the observational constraints on the 
luminous mass are considered, the satellite model could represent a DM 
dominated DSph.
For these models, we have assumed 
$(M/L)_{\mbox{real}}=10$ M$_{\odot}/$L$_{\odot}$
for the calculations of surface brightness, which is the lower limit of the 
mass-to-luminosity ratio for DM dominated satellites. A higher
$(M/L)_{\mbox{real}}$ would represent fainter satellite galaxies (for the same
DM content) and an apparently faster dissolution process. 

\begin{table}
\caption{Physical parameters of the isolated King's model of Sgr (f-models):
core radius, $r_o$, central velocity dispersion, $\sigma_o$, 
total mass, $M_t$, dimensionless potential, $W_o$ and tidal radius, $r_t$}
\label{tbl1}
\begin{tabular}{lllll}
\hline
$r_o$ & $\sigma_o$ & $M_t$ & $W_o$ & $r_t$ \\
 (kpc) &  (km/s) & ($10^7$ M$_{\odot}$)& & (kpc) \\
\hline

0.527 & 15.0 & 12.0 & 3.26 & 2.736\\
\hline
\end{tabular}
\end{table}

The initial apocenter and pericenter and the period 
of the low 
eccentricity orbit (f-A) and the high eccentricity orbits (f-B1 and f-B2) 
are listed in Table \ref{tbl1-2}.

\begin{table}
\caption{Parameters of the orbits of the f-models}
\label{tbl1-2}
\begin{tabular}{llll}
\hline
Model & $r_{min}$ & $r_{max}$ & Period\\
      &  (kpc)    &  (kpc)    & (Gyr)\\
\hline
f-A   & 12        & 18        & 0.23\\
f-B1  & 8         & 38        & 0.45\\
f-B2  & 10        & 70        & 0.95\\
\hline
\end{tabular}
\end{table}

\subsection{s-models}

 In the other possible scenario we assume that either the satellite 
has been formed inside the tidal tail of a major merger (e.g. 
numerical simulations by Barnes \& Hernquist 1992, and
the observational counterpart by Duc \& Mirabel
\cite{Duc97}) or it has been slowly accreted. 

 In the first case, if the dwarf galaxy is formed in equilibrium with the tidal 
force of the environment, it does not contain a significant amount of 
DM (Barnes \& Hernquist \cite{Barnes92}).

\begin{table}
\caption{Galaxy parameters of the s-model satellites}
\label{tbl2}
\begin{tabular}{llllll}
\\
\hline
Model&$r_o$&$\sigma_o$&     $M_t$          &$W_o$&$r_t$ \\
      &(kpc)& (km/s)  &($10^7$ M$_{\odot}$)&     & (kpc)  \\
\hline
s-A  & 0.06 & 11.0     &    0.93      &3.78 & 0.55 \\
s-B1 & 0.1  & 11.0     &    1.66      &3.95 & 0.99 \\
s-B2a& 0.1  & 15.0     &    5.27      &5.42 & 1.93 \\
s-B2b& 0.1  & 11.0     &    2.33      &4.87 & 1.48 \\
s-B2c& 0.3  & 15.0     &    6.04      &2.98 & 2.02 \\
\hline
\end{tabular}
\end{table}

\begin{table}
\caption{Orbital parameters of the s-model satellites} 
\label{tbl2-2}
\begin{tabular}{lllll}
\\
\hline
Model&$r_{min}$&$r_{max}$&$r_{ave}$ & Period\\
      &  (kpc)  & (kpc)  &  (kpc)   &  (Gyr)\\
\hline
s-A  & 15      & 20      &   20     &  0.25\\
s-B1 &  8      & 40      &   35     & 0.45\\
s-B2 & 10      & 74      &   50     & 1.25\\
\hline
\end{tabular}
\end{table}

 In the second case, according to cosmological models of hierarchical structure
formation, satellite systems are produced around massive galaxies.
These satellites could contain DM halos (e.g. Cole et al. 
\cite{Cole94}).
Tidal forces could be negligible in the outer regions of the galaxy where the
satellite is formed.  However, if we assume that 
the satellite does not go through the denser and central regions of the 
main galaxy in the first perigalacticon, 
we could expect that an initially massive satellite 
falls slowly on the center of the galaxy by dynamical friction, loses
part of its mass and reaches a central orbit. In this way, the
satellite has time to modify its internal structure and to reach 
the equilibrium with the environment.

 DSphs described in the two last scenarios have to be in equilibrium with the 
tidal forces of the environment. In the paper by G\'omez-Flechoso
\& Dom\'{\i}nguez-Tenreiro (\cite{Gomez97}), the 
structural parameters (total mass, $M_t$, velocity dispersion, 
$\sigma_o$, core radius, $r_o$, etc.) of a satellite in the tidal 
field of the primary galaxy have been estimated. Those authors have
proved that, in general, a galaxy in a tidal field can be described by
a two parameter distribution function. 
They have solved altogether the Poisson equation and the
collisionless Boltzmann equation for a galaxy, taking 
into account the potential of the galaxy
and the tidal potential of the environment. Only spherical terms of the
tidal field were considered in this theory.
However, the obtained {\it equilibrium} solutions are better representations 
of the system than {\it isolated} models.

 This result suggests that in our problem we could represent a DSph 
in equilibrium with the tidal field 
of the primary galaxy as a {\it modified King's sphere} (see G\'omez-Flechoso
\& Dom\'{\i}nguez-Tenreiro \cite{Gomez97}) with two free parameters. 
 We have chosen the central velocity dispersion,
$\sigma_o$, and the core radius, $r_o$, as input parameters, because  
both can be determined from observations. 
In our simulations, the values for these two parameters 
are $r_o \sim 0.06-0.3$~kpc
and $\sigma_o \sim 11-15$~km/s, which reproduce the characteristics
of the Milky Way satellites. Thus, we will try to explain the 
Sagittarius satellite as a typical DSph which has evolved orbiting
for a long time in the Galaxy potential.
The other parameters of the model (total mass, tidal radius and
dimensionless central potential) are automatically determined
by the tidal potential at each position of the orbit. 

 The satellite parameters of the models which have been performed 
are listed in Table \ref{tbl2}. 
The second column
is the core radius, $r_o$, and the third column is the central velocity
dispersion, $\sigma_o$, which are input parameters of the modified
King's spheres.
The total mass, $M_t$, the dimensionless central potential, $W_o$, and the
tidal radius, $r_t$, (columns 4, 5 and 6) are output parameters obtained 
by solving the collisionless Boltzmann equation with the tidal potential
at the averaged distance of the orbit to the Galaxy center 
(parameter $r_{ave}$ in Table \ref{tbl2-2}).

 As it can be seen in Table \ref{tbl2}, the dimensionless central potential 
and the mass of the satellite decrease for inner positions of the 
equilibrium satellite for models with the same $r_o$ and $\sigma_o$. 
Furthermore, the mass of the models is smaller 
than the mass inferred from observations using kinematic arguments 
(Ibata et al. \cite{Ibata97}, Mateo \cite{Mateo94}) and it is in agreement with the observed
luminous mass, assuming $(M/L)_{\mbox{real}} \sim 2-5$ M$_{\odot}/$L$_{\odot}$. 
Therefore, we have assumed 
$(M/L)_{\mbox{real}}=2$ M$_{\odot}/$L$_{\odot}$ for all
the s-models, that is a typical value for the 
stellar population of a DSph. The low value of the mass-to-luminosity
ratio is in agreement with satellites either formed in tidal tails of major 
accretion events or tidally modified by orbiting for a long time in a tidal 
potential.

The apocenter, $r_{max}$, the pericenter, $r_{min}$, and the period 
of the
orbits are listed in Table \ref{tbl2-2}.
The s-A orbit is an orbit of
low eccentricity in the inner region of the Galaxy and s-B1 and s-B2 orbits 
have higher eccentricity.

\section{Numerical details}
\label{numerical}

 The models have been evolved using the treecode algorithm kindly provided
by Barnes \& Hut (\cite{Barnes86}) with a tolerance parameter $\theta=0.7$ and 
a time-step $\Delta t = 1$ Myr.
 The number of particles of the luminous and dark halo components of 
the primary galaxy are 15671 and 29648, respectively, and the mass
of the dark matter particles is 3 times larger that for the
luminous particles. 
 All the satellite models have 4000 equal-mass particles, except s-B2a 
and s-B2c which have 8000 particles.

We use a softening length varying 
proportionally
to the cubic root of the particle mass of the component, in order
to avoid well-known usual 
numerical effects in the simulations (e.g. Merritt \cite{Merritt96}; Theis
\& Spurzem \cite{Theis99}). For 
the luminous particles of the Galaxy it is $\epsilon_L =0.23$ kpc and 
for the dark matter halo  $\epsilon_{DH} = 0.33$ kpc. The f-models
have $\epsilon_S=0.06$ kpc, but this value is changed to
$\epsilon_S=0.05$ kpc for the s-A, s-B1 and s-B2b satellites and 
$\epsilon_S=0.04$ kpc for the s-B2a and s-B2c satellites.

\section{Results}

\subsection{The main galaxy}

The main galaxy develops a bar-like structure,
described by Fux (\cite{Fux97}) for an isolated model of our 
Galaxy. 

The global effects of the satellite on the primary
galaxy are weak, since the mass ratio of both objects is huge. Besides,
the poor resolution of the Galaxy model prevents 
a detailed description of
the local effects of the Sagittarius accretion on the Milky Way. 
Therefore, we only deal with dynamical effects felt by the satellite galaxy.

\subsection{f-models}
\label{fastsat}

For our f-models, we begin 
the simulations when the satellite has already reached the inner 
regions of the primary galaxy.
 A DSph galaxy on a low eccentricity orbit 
in these inner dense regions of the Galaxy 
undergoes
stronger disruptions than on more eccentric orbits, 
because the tidal forces are stronger
at all positions on the trajectory. 

\subsubsection{f-A satellite} 

\begin{figure}
 \resizebox{\hsize}{!}{\includegraphics{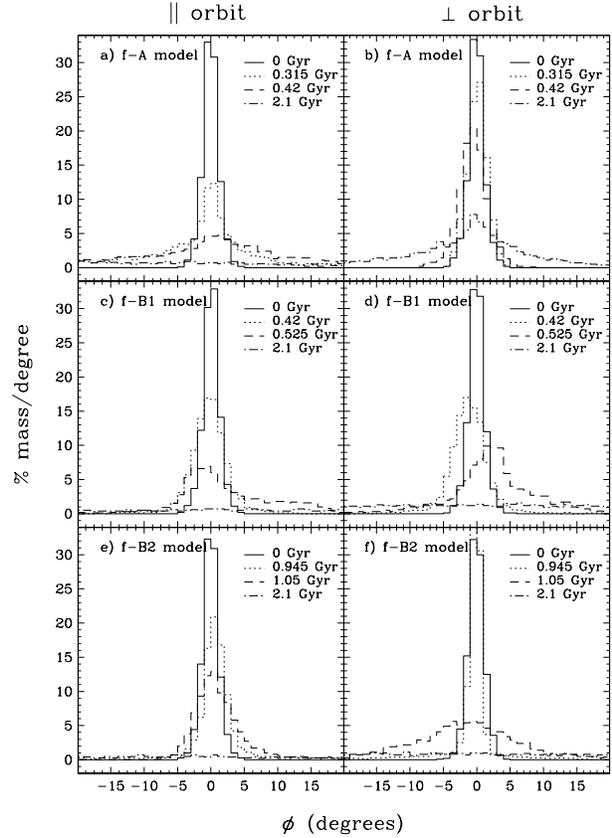}}
  \caption{f-models: mass distribution parallel 
   ({\bf a}, {\bf c} and {\bf e}) and perpendicular 
   ({\bf b}, {\bf d} and {\bf f}) to the
   orbit as function of the angle $\phi$ to the satellite center, 
   as seen from 24 kpc away, which is the present distance between 
   the Sun and the Sagittarius DSph}
  \label{f2}
\end{figure}

\begin{figure}
 \resizebox{\hsize}{!}{\includegraphics{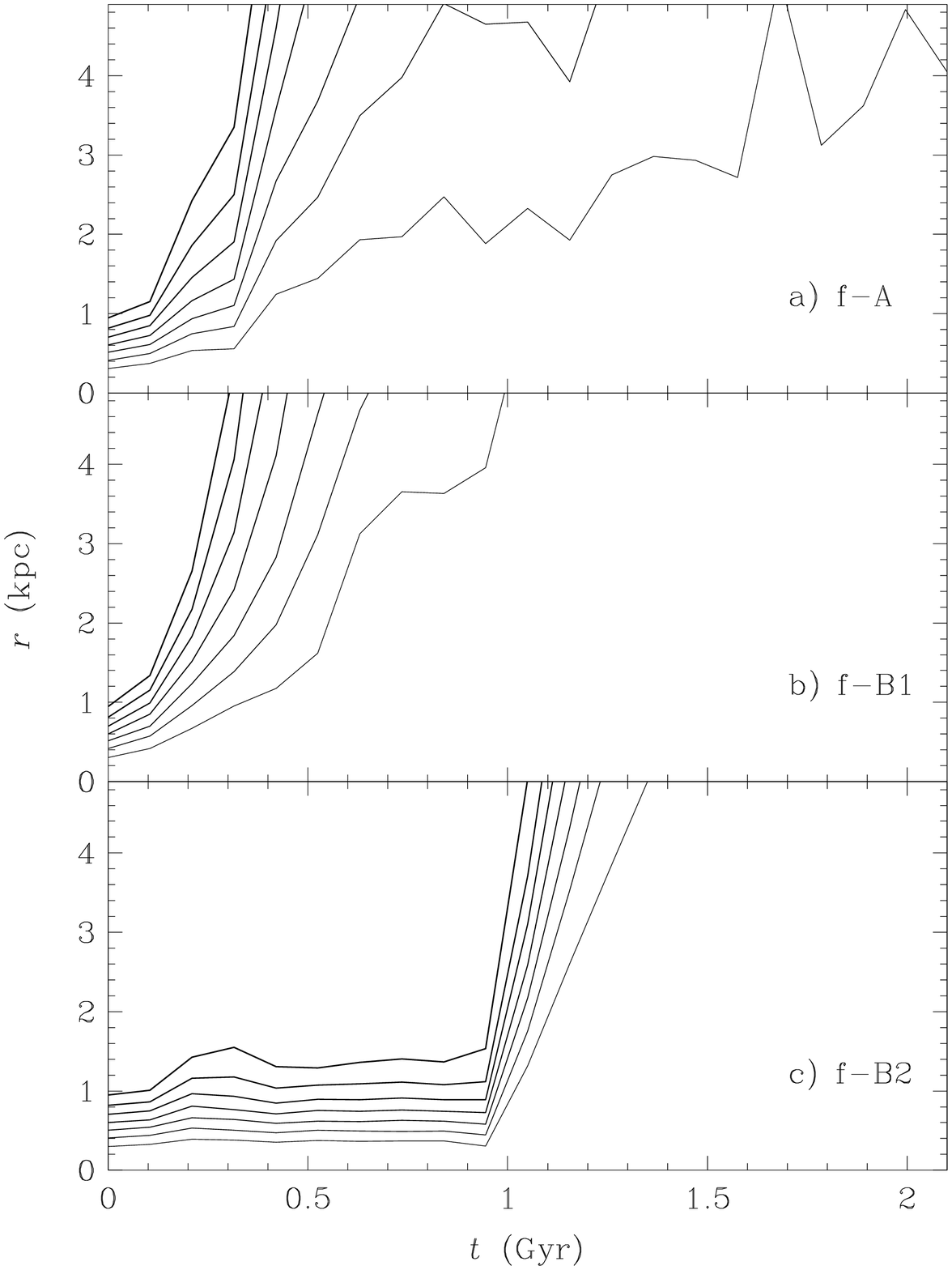}}
  \caption{Radii which enclose 70\%, 60\%, 50\%, 40\%, 30\%, 20\% and 10\% 
          of the initial mass of the satellite, for f-models}
  \label{f3}
\end{figure}

In Fig. \ref{f2}a (b), we plot the angular distribution of the 
f-A satellite along (perpendicular to) 
the orbit (which has eccentricity $e=0.2$) 
as seen from 24 kpc away (it corresponds
to the Solar neighbourhood viewpoint), for four snapshots.
In this case, the lifetime of the dwarf galaxy, 
before significant 
disruption, is nearly 0.4 Gyrs 
(Fig. \ref{f3}a). 

\begin{figure}
 \resizebox{\hsize}{!}{\includegraphics{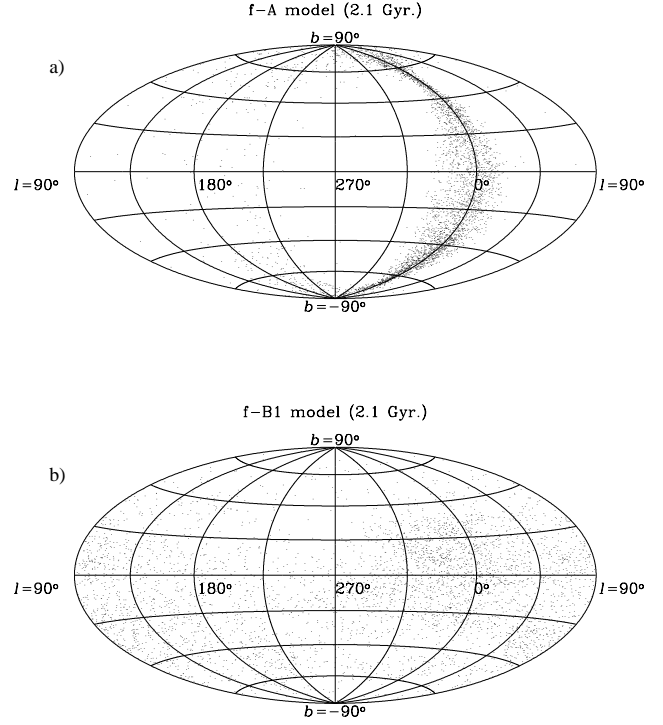}}
 \caption{Aitoff projection of the f-A and f-B1 models at the 
          end of the simulation (2.1 Gyr)}
 \label{f4}
\end{figure}

\begin{figure}
 \resizebox{\hsize}{!}{\includegraphics{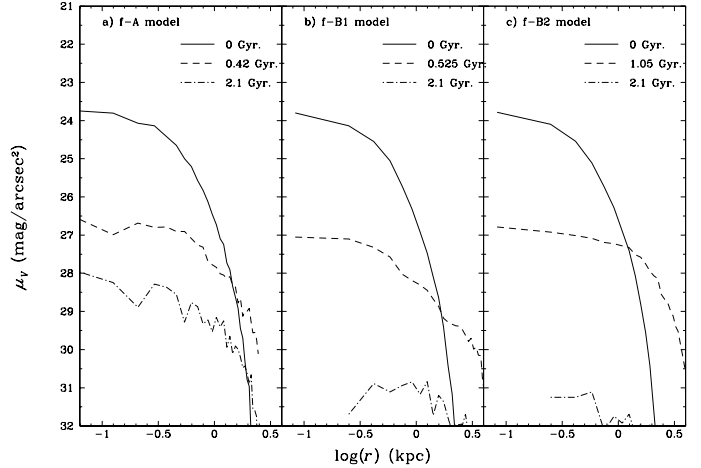}}
  \caption{Surface brightness $\mu_V$ of the f-models, assuming
   $(M/L)_{\mbox{real},V}=10$ M$_{\odot}/$L$_{\odot}$}
  \label{f5}
\end{figure}

 The material from the satellite galaxy is tidally stripped, forming
a long stream and then, 
after 1.2 Gyrs, an almost close great circle in an Aitoff projection,
which survives for a long time (Fig. \ref{f4}a).
Moreover, as we can see in Fig. \ref{f2}b, 
the mean width perpendicular to the orbit 
on the sky is $7^o$ (it was only $1.9^o$ at the initial time) 
and the projected surface brightness (Fig. \ref{f5}a) is 
5 mag fainter at the end of the simulation (after 2.1 Gyrs).

\subsubsection{f-B1 and f-B2 satellites} 

 For both f-B1 and f-B2 orbits ($e=0.64$ and 0.75, respectively), 
disruption mainly occurs at perigalacticon, 
because
tidal forces are more efficient at small galactocentric radii. 
The lifetime of a satellite depends strongly on the orbit shape.
The f-B1 satellite is destroyed after 0.5 Gyr (Fig. 
\ref{f3}b) whereas 
the f-B2 satellite survives for 1 Gyr (Fig. \ref{f3}c) due to the longer 
period of its orbit, at this time the f-B2 satellite undergoes a close 
{\it interaction} ($r \sim 2$ kpc) 
with the center of the primary 
galaxy and it is tidally destroyed.

 The final destruction of dwarf galaxies is more efficient in 
our more 
eccentric orbits and they present fainter surface 
brightness at the end
of the simulation than the satellite on a low eccentric orbit (Fig. 
\ref{f5}). The satellite particles of f-B models are spread 
over all directions and no predominant streams are formed,
as it can be seen in Fig. \ref{f4}b for the f-B1 model.
In Figs.
\ref{f2}d and \ref{f2}f, 
we have represented the width (perpendicular to the orbit) 
of the stream. At the end of the simulations, the f-B2 models
do not have any predominant peak in the  
mass distribution perpendicularly to the initial orbit.

\begin{figure}
 \resizebox{\hsize}{!}{\includegraphics{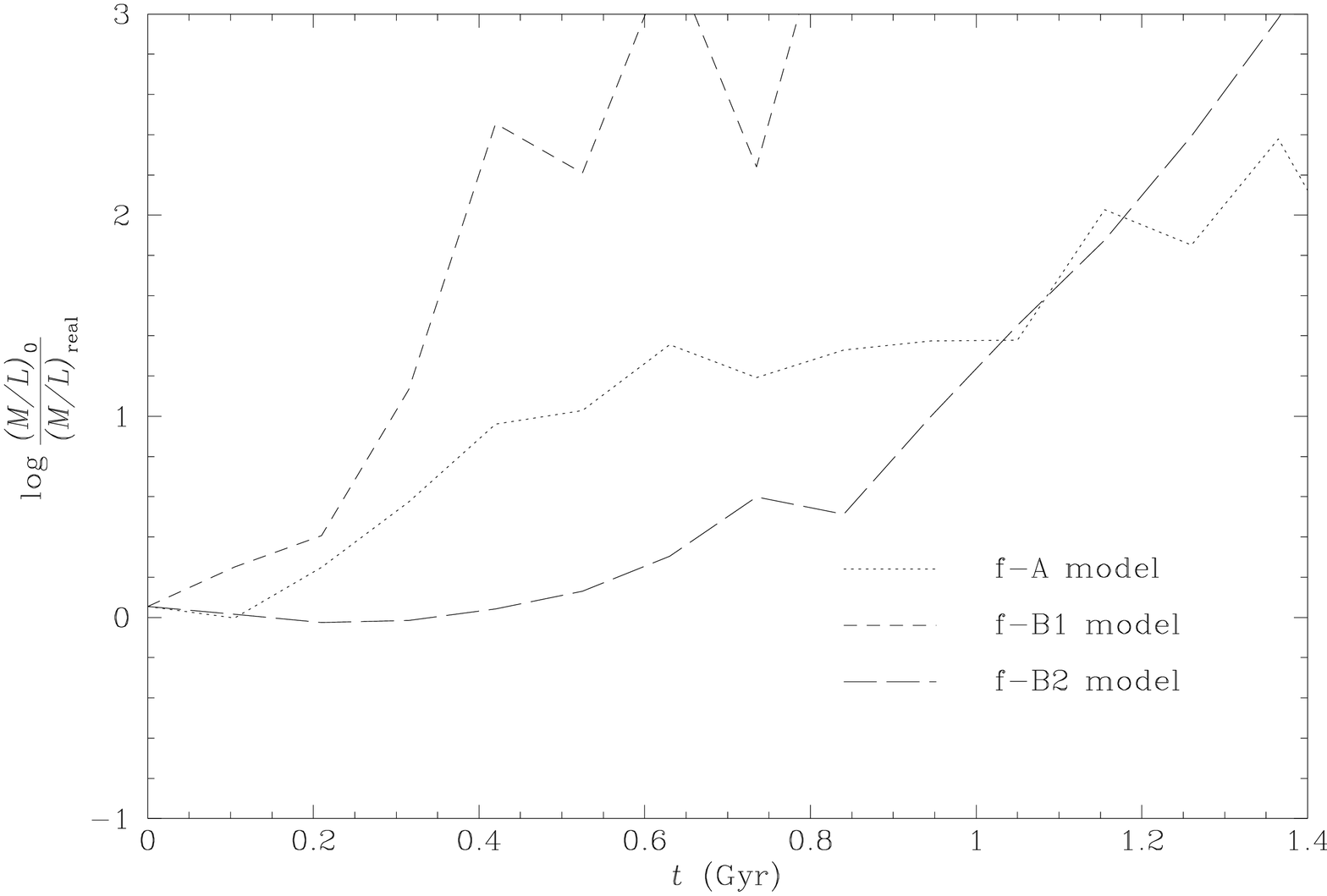}}
  \caption{Evolution of the log of 
   the ratio between the mass-to-luminosity 
   relation, $(M/L)_o$, inferred from the velocity dispersion values
   under dynamical equilibrium conditions, 
   and the real mass-to-luminosity relation, $(M/L)_{\mbox{real}}$,
   for the f-models}
  \label{f6}
\end{figure}

 As a general result, due to the tidal field on the satellite
we observe: 
i) a modification of the internal structure of the satellite for
both cases of orbits 
(eccentric and quasi-circular), with increase of the 
projected velocity dispersion, and ii) a 
continuous loss of satellite mass and luminosity.
Anyone tempted to infer a $(M/L)_o$ ratio from such experiments
by assuming virial conditions would find values  $100-200$ times
higher than the true ones at the disruption time (Fig. \ref{f6}),
confirming the results of K97.

\subsection{s-models}
\label{slowsat}

 In this subsection, we analyze the interaction 
effects on a DSph in equilibrium with 
the galaxy potential. Either it could have fallen down slowly from an intermediate 
region
in the denser parts of the primary galaxy, loosing part of its mass and 
becoming a low DM satellite, or it could have been formed in a tidal 
tail of a major accretion event. 

\subsubsection{s-A model}

\begin{figure}
 \resizebox{\hsize}{!}{\includegraphics{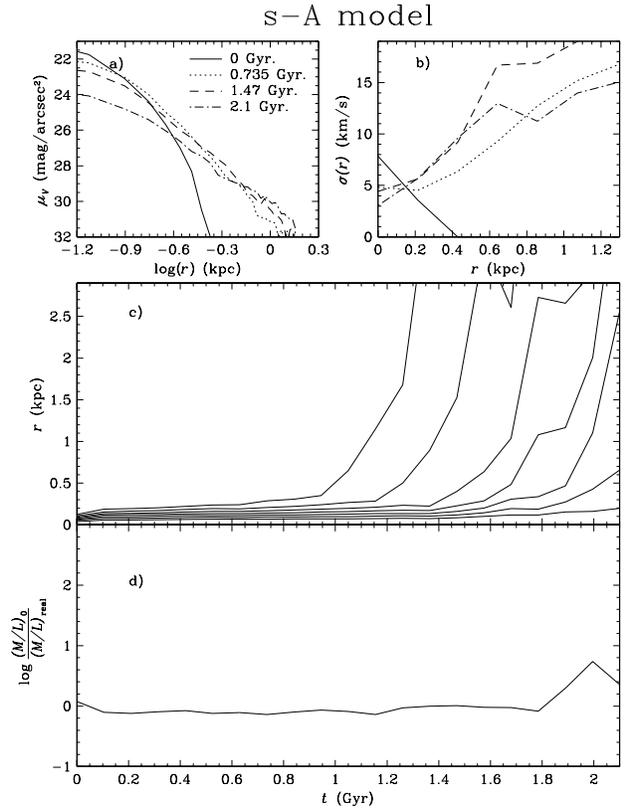}}
  \caption{s-A model: 
    {\bf a} surface brightness $\mu_V$ assuming $(M/L)_{\mbox{real},V}=2$ 
    M$_{\odot}/$L$_{\odot}$, 
    {\bf b} velocity dispersion of the satellite as a 
    function of the radius as same timesteps as in a), 
    {\bf c} evolution of the radii which enclose from 70 to 10 \% of mass 
    of the satellite from the top to the bottom,
    and {\bf d} evolution of the mass-to-luminosity ratio 
    measured from the velocity dispersion. The observational
    values of $\mu$, $\sigma$ are given in Table \ref{Sgr_tbl}}
  \label{f7}
\end{figure}

\begin{figure}
 \resizebox{\hsize}{!}{\includegraphics{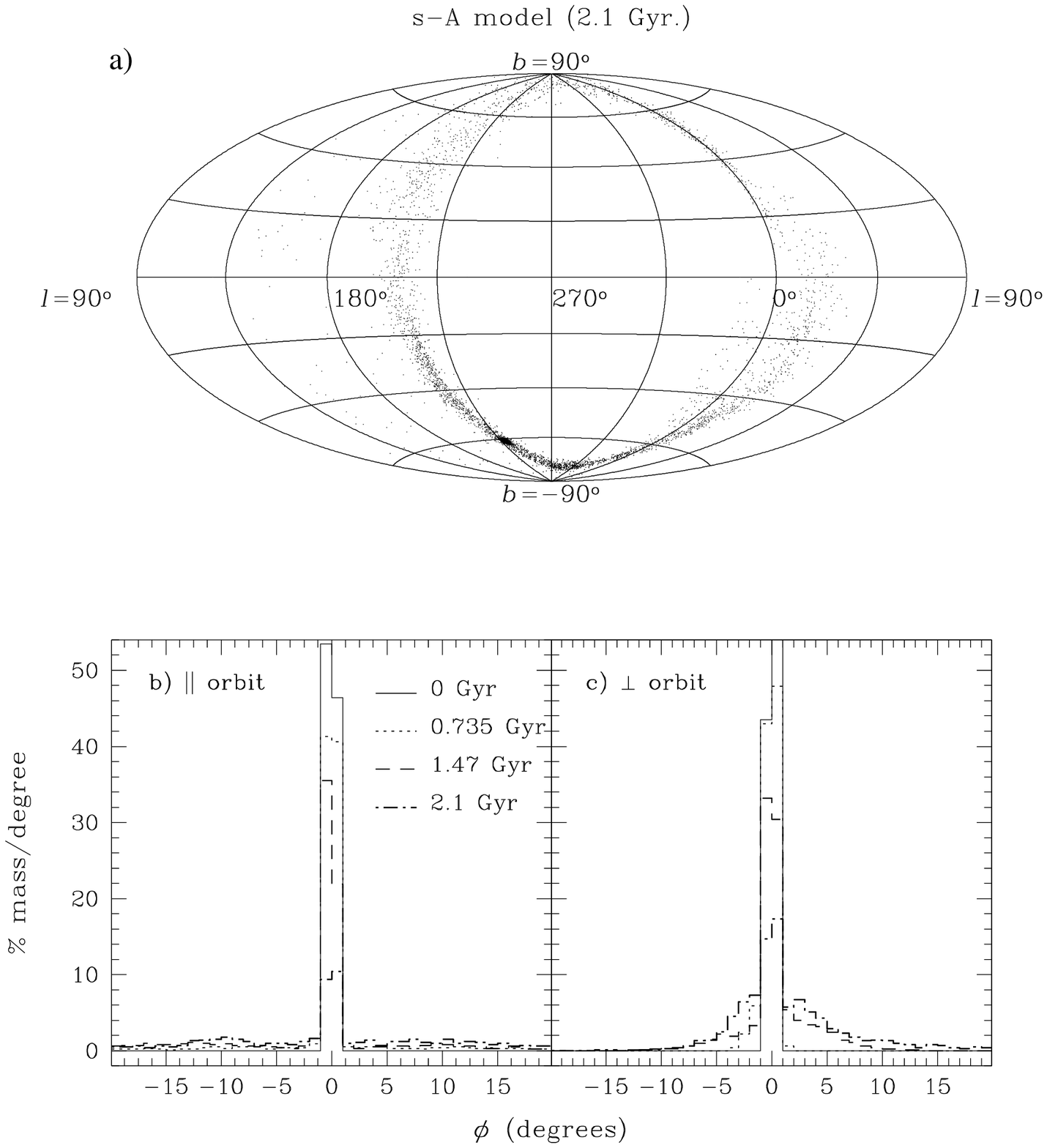}}
 \caption{{\bf a} Aitoff projection of the s-A model at the end of the
            simulation (2.1 Gyr), and mass distribution 
          {\bf b} parallel and {\bf c} perpendicular to the orbit as 
            function of the angle $\phi$ to the satellite center}
\label{f8}
\end{figure}

 DSphs which are theoretical 
equilibrium solutions to Sgr at a circular inner orbit
are small and low mass galaxies. The tidal field is almost constant
along this quasi-circular orbit and, therefore, interaction 
effects on the satellites
in equilibrium are not important. For the s-A orbit, there 
is a continuous loss of mass, but the satellite has still
$\sim 20 \%$ of the initial mass (Fig. \ref{f7}c) 
after 2 Gyrs 
and it is still detectable (Fig.
\ref{f7}a). The central surface brightness has 
only changed by 2.5 mag (Fig. \ref{f7}a), 
evolving from 21.3 mag arcsec$^{-2}$ to 24.8 mag arcsec$^{-2}$.
The line-of-sight velocity dispersion grows in the outer parts of the
satellite in agreement with findings by K97, 
remaining almost constant in average in the inner parts ($<0.2-0.3$ kpc) 
(Fig. \ref{f7}b). 
We have 
calculated the 
mass inferred from the line-of-sight velocity dispersion, measured inside a 
radius of 
0.5 kpc from the satellite center, using Eq. (\ref{ML}). 
Since the central velocity dispersion and the 
projected density (Fig. \ref{f7}a) 
evolve slowly in the central region of
the satellite at the beginning of the simulation, the inferred 
mass-to-luminosity ratio
remains almost constant, contrary to the f-models. However, as 
we have explained above, the real mass of the satellite decreases and 
it is only $\sim 20 \%$ of the initial value after 2 Gyrs. 
The structural evolution is slow (Fig.~\ref{f7}c) 
and it leads to a lower central surface
brightness at the end of the simulation and a $(M/L)_o$ ratio calculated from
the velocity dispersion (inside 0.5 kpc) which is 5 times higher than 
the real one (Fig.~\ref{f7}d). This effect
could be dramatically increased if the velocity dispersion is measured inside a
larger radius, because of the 
velocity dispersion increase in the outer 
region of the satellite, leading to a 
calculated $(M/L)_o$ ratio up to 10 times larger than the real one. 
Very faint tidal streams are formed, which
are spread along $\sim 75-100 \%$ of the orbit with a spatial width $\sim
6^o$ (Fig. \ref{f8}). 

\subsubsection{s-B models}

\begin{figure}
 \resizebox{\hsize}{!}{\includegraphics{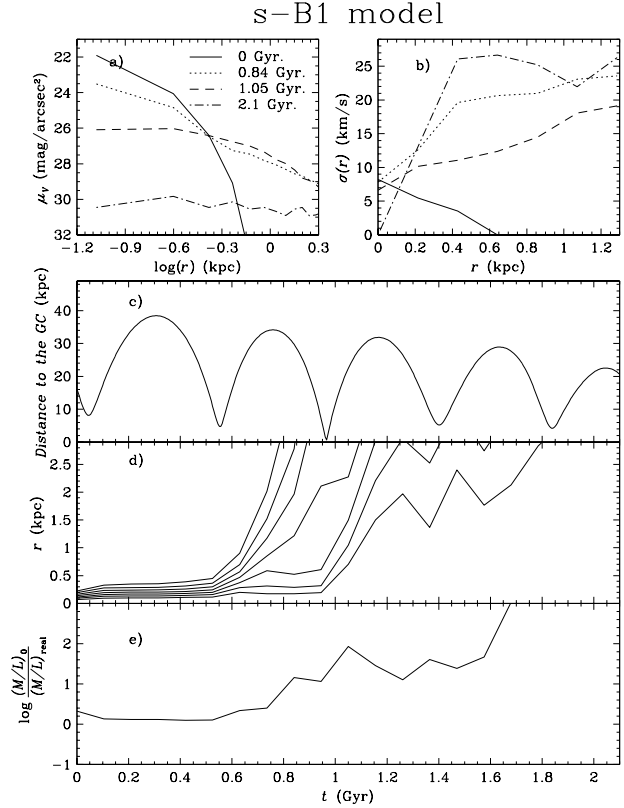}}
  \caption{s-B1 model: 
    {\bf a} surface brightness $\mu_V$ assuming $(M/L)_{\mbox{real},V}=2$ 
    M$_{\odot}/$L$_{\odot}$, 
    {\bf b} velocity dispersion of the satellite as a function of the radius,
    {\bf c} distance to the satellite the Galactic Center, 
    {\bf d} evolution of the radii which enclose 70, 60, ..., 10\% of 
    the satellite mass (lines from left to right),
    and {\bf e} evolution of the mass-to-luminosity ratio 
    measured from the velocity dispersion. The observational values
    of $\mu$ and $\sigma$ are given in Table \ref{Sgr_tbl}}
  \label{f9}
\end{figure}

 Perigalacticon and apogalacticon of high eccentric
s-B1 and s-B2 orbits decrease with the time.
The most dramatic example is the {\bf s-B1 orbit}. The DSph passes 1 kpc 
from 
the galaxy center at its third passage at perigalacticon 
(Fig. \ref{f9}c), because of the 
orbital energy loss. During this passage
the satellite is strongly tidally stripped and it loses 
most of its mass (Fig. \ref{f9}d). 
The surface brightness of the satellite 
decreases with time (Fig. \ref{f9}a)
and the velocity dispersion 
and the mass-to-luminosity ratio increase in the outer parts
(Figs. \ref{f9}b and \ref{f9}e). 
The satellite is finally completely destroyed after nearly 1 Gyr. 

 s-B2 is 
the most external and eccentric orbit which we have chosen.
As the other models, it 
fulfills at certain times the position and velocity 
constraints for the Sagittarius DSph orbit in
the observational range (Ibata et al. \cite{Ibata97}).

\begin{figure}
 \resizebox{\hsize}{!}{\includegraphics{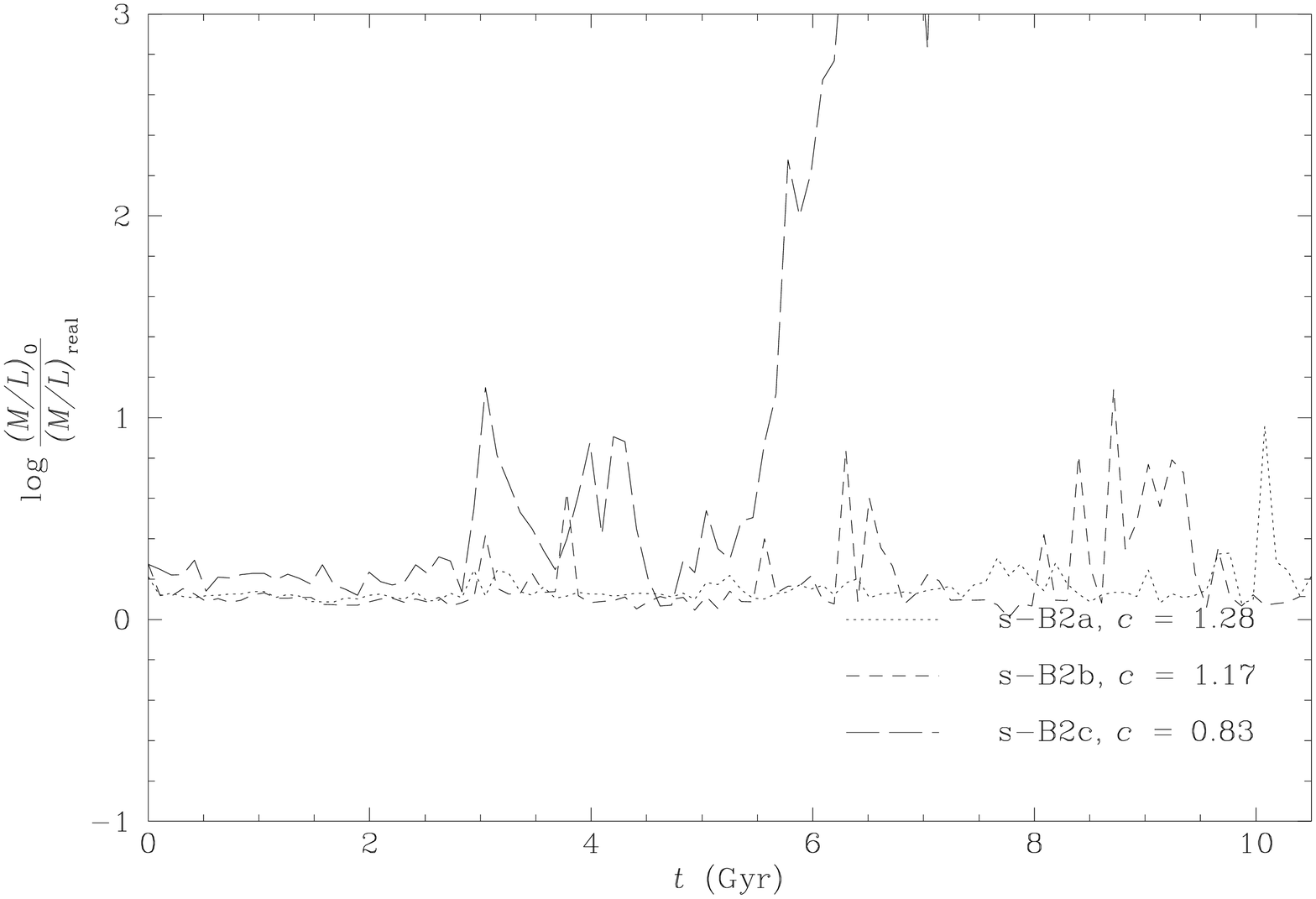}}
  \caption{Evolution of the log of 
  the ratio between the mass-to-luminosity relation,
  $(M/L)_o$, inferred from
   the velocity dispersion values, and the real mass-to-luminosity 
   relation, $(M/L)_{\mbox{real}}$, for s-B2 satellites. The parameter 
   $c=\log(r_t/r_o)$ is the concentration of the model}
  \label{f10}
\end{figure}

 In order to investigate this case in more details, 
we have selected three different satellite models for orbiting on this
trajectory (s-B2a, 
s-B2b and s-B2c). The concentration (defined as $c=\log(r_t/r_o)$)
varies from 1.28 (s-B2a) to 0.83
(s-B2c). The concentration determines the fate of the satellite, as well
as the evolution of the $(M/L)_o$ ratio. The more concentrated the satellite is, 
the less variation of the inferred $(M/L)_o$ ratio it suffers (Fig. \ref{f10}).

\begin{figure}
 \resizebox{\hsize}{!}{\includegraphics{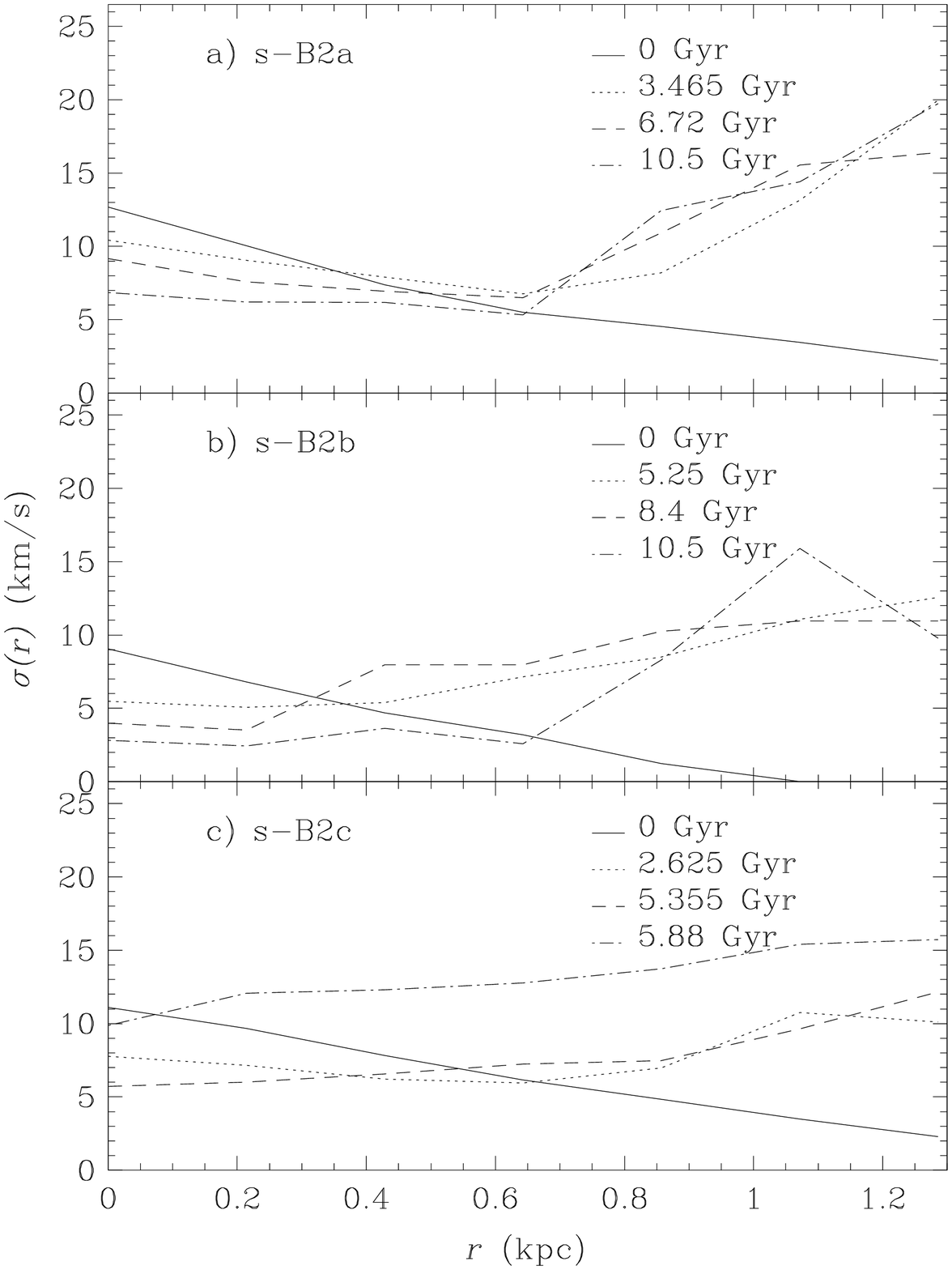}}
  \caption{Velocity dispersion, $\sigma(r)$, at four snapshots, for 
           s-B2 satellites}
  \label{f11}
\end{figure}

 The $(M/L)_o$ ratio in Fig. \ref{f10} has been calculated for 
the central region of the satellite ($< 0.5$ kpc), 
where the tidal effects on the mass and the velocity dispersion 
are weaker. However, even in the most concentrated satellite (s-B2a), we
obtain $(M/L)_o$ ratios which are $\sim 10 (M/L)_{\mbox{real}}$, these values rise
up to several hundreds for the least concentrated model (s-B2c). 

 It is interesting to notice the strong variations of the 
inferred $(M/L)_o$ ratio with time. This behaviour is 
caused by the evolution of the surface mass density 
when the satellite suffers a close encounter with the core of
the primary galaxy. It leads to a successive rearrangement of the 
internal structure of the satellite.
Another important parameter in the $(M/L)_o$ calculations is the angle between 
the observer-satellite line 
and the main axis of the satellite. The strong 
anisotropies of the satellite (tails along the orbit, anisotropic
velocity dispersion, etc) could produce various $(M/L)_o$ values (as already 
suggested
by K97 and KK98).

 The line-of-sight velocity dispersion evolves as shown in 
Fig.~\ref{f11}. 
The projected velocity dispersion decreases in the central part of the satellite
($r<0.5$ kpc) and it increases for $r>0.5$ kpc. Thus,
the $(M/L)_o$ ratio calculated from Eq.~(\ref{ML})
varies, depending on the limit radius used for the 
measurement 
of the {\it central} velocity dispersion
(this limit radius is related to the observational resolution).

\begin{figure}
 \resizebox{\hsize}{!}{\includegraphics{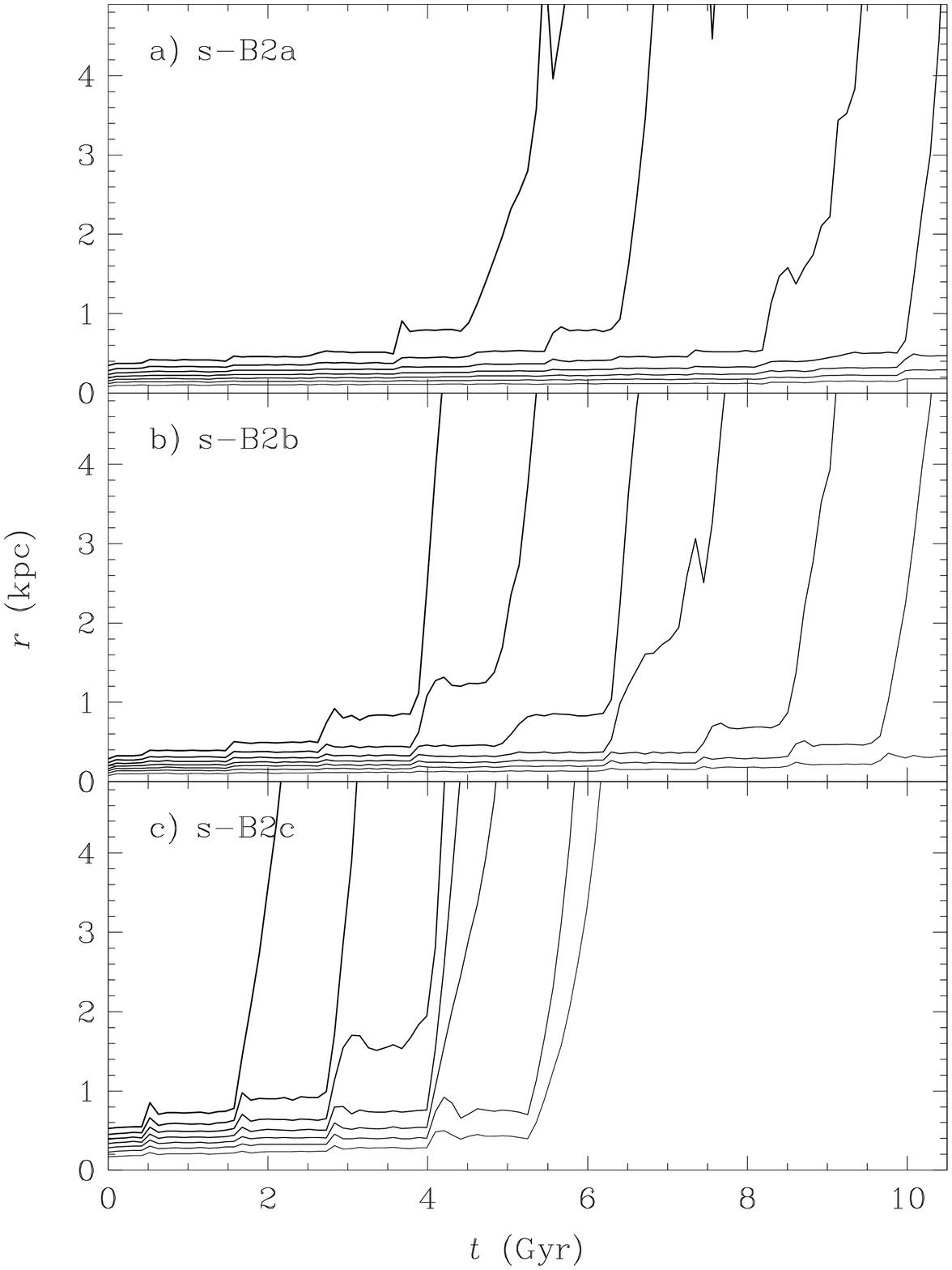}}
  \caption{Radii which enclose 70\%, 60\%, 50\%, 40\%, 30\%, 20\% and 10\% 
          of the initial mass of the satellite, for s-B2 models}
  \label{f12}
\end{figure}

{\it s-B2a model:} 

\begin{figure}
 \resizebox{\hsize}{!}{\includegraphics{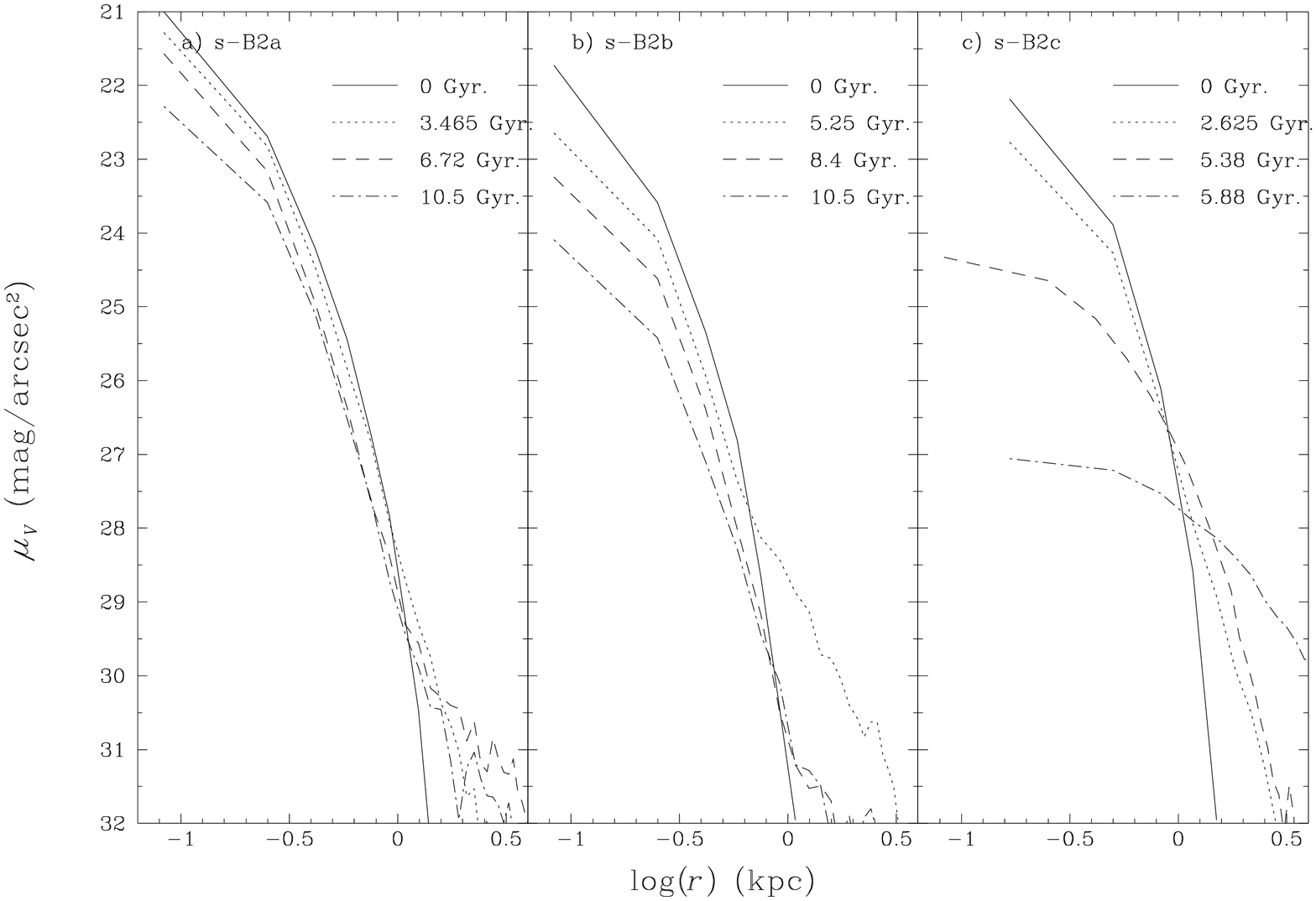}}
  \caption{Surface brightness $\mu_V$ of the s-B2 satellites, assuming
   $(M/L)_{\mbox{real},V}=2$ M$_{\odot}/$L$_{\odot}$}
  \label{f13}
\end{figure}

 The {\bf s-B2a model} undergoes a long term evolution. It is the most 
concentrated model that we have chosen and it survives 
for at least 10 Gyrs. A mild loss of mass is produced, mainly 
at perigalacticon:
the satellite looses the outer layers of mass, but a core enclosing 30\% of 
the initial mass subsists (Fig. \ref{f12}a). The limit radius is smaller 
than the 
initial one, but the half-brightness radius increases (the final
satellite is less concentrated). The central surface brightness decreases
(Fig. \ref{f13}a), changing from $\sim 21.0$ mag arcsec$^2$ to $\sim 22.0$ 
mag arcsec$^2$ (assuming $(M/L)_{\mbox{real}}=2$ M$_{\odot}/$L$_{\odot}$).
The final s-B2a satellite is smaller ($r_{hb}=0.13$ kpc)
than the observed Sgr DSph ($r_{hb}=0.55$ kpc) and
the $(M/L)_o$ ratio is not large enough to reproduce the inferred $(M/L)_o
= 50$ M$_{\odot}/$L$_{\odot}$ by Ibata et al. (\cite{Ibata97}).

{\it s-B2b model:} 

 The mass loss of the {\bf s-B2b satellite}
(Fig. \ref{f12}b) is stronger than that of the s-B2a satellite. 
The particles of the satellite outer region 
are stripped, mainly 
at perigalacticon. The 
tidal stripped material develops a stream forward and backward from 
the satellite along the orbit.
In Fig. \ref{f13}b, a low surface brightness tidal extension of 
the satellite can be seen
at intermediate epochs. At the final snapshot ($t=10.5$ Gyr), the residual 
core of the satellite is still detectable, it has $\sim 15 \%$ of the initial 
mass (Fig. \ref{f12}b) and a central 
projected surface brightness $\mu_o \sim 23.8$ 
mag arcsec$^{-2}$ assuming 
$(M/L)_{\mbox{real}}=2$ M$_{\odot}/$L$_{\odot}$. The initial central surface brightness for the
same $(M/L)_{\mbox{real}}$ ratio was $21.3$ mag arcsec$^{-2}$. The stream formed close to the 
satellite has a surface brightness which is 5.0-6.5 mag fainter than the center
of the satellite. This stream is similar to those formed by extra-tidal stars
observed close to some DSph satellites.

\begin{figure}
 \resizebox{\hsize}{!}{\includegraphics{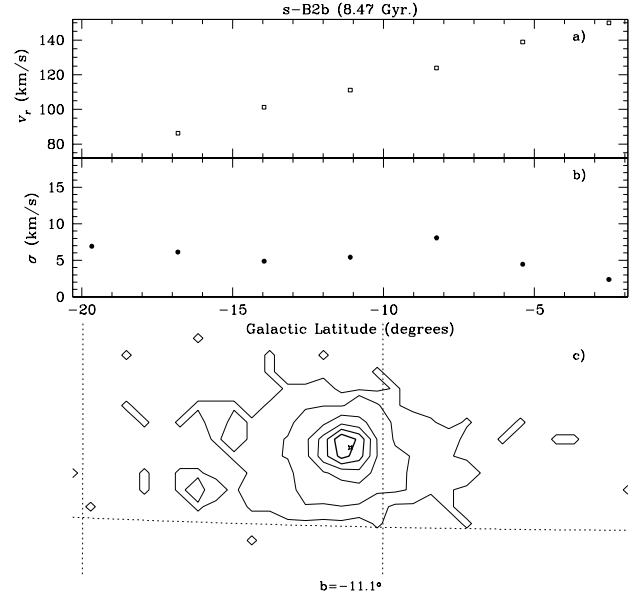}}
  \caption{s-B2b satellite, at 8.47 Gyr: 
           {\bf a} Radial velocity, $v_r$, along the orbit
	   (constant galactic latitude),
           {\bf b} velocity dispersion, $\sigma$, along the orbit 
           and 
	   {\bf c} contour density map at the same scale as the a) 
           and b) plots 
           (the star represents the mass 
           center of the satellite, located at $b=-11.1^o$), 
           the 
           contours correspond to $\mu=$ 23.2, 23.6, 24.0, 24.7, 26.8 and 
           28.7~mag arcsec$^2$ from the center}
  \label{f14}
\end{figure}

 At some snapshots (i.e. 3.7 Gyr, 4.83 Gyr, 6.2 Gyr, 7.3 Gyr, 8.45 Gyr
and 9.6 Gyr), when the tidal streams of the satellite are already formed,
the radial velocity, $v_r$, 
and the gradient of the radial velocity, $|dv/db|$,
of the s-B2b satellite and of the
stream around the satellite reproduce the observations of Sgr 
DSph region. The observational values of $v_r$ and $|dv/db|$ of Sgr
are given in Table \ref{Sgr_tbl}.
However, the values of both quantities in the models are strongly dependent on
the position and orientation of the orbit. 
Therefore, small perturbations
in the orbit of the models 
could lead to differences in $v_r$ of several 
10 km s$^{-1}$. 
In general, the sign of the variation of the radial velocity depends 
on the position of the satellite along the orbit. For some particular 
positions, the radial velocity gradient is almost 0 km s$^{-1}$/degree. 
The s-B2b satellite at time 8.47 Gyr is an example of a model which has a
good agreement in position with the observations of Sgr DSph
(Fig. \ref{f14}c).
From the kinematical point of view, the particles in the stream 
show a velocity gradient $|dv/db| \sim
4$ km s$^{-1}$/degree (Fig. \ref{f14}a), which is 
slightly higher than the observed value ($|dv/db| =3$ km s$^{-1}$/degree), 
whereas the radial velocity is slightly lower.
Moreover, this satellite model presents a velocity dispersion ($\sigma 
\sim 6$ km/s, Fig. \ref{f14}b) lower 
than the 
observed value ($\sigma \sim 11.4$ km/s). However,
the projected velocity dispersion of the stream remains almost 
constant and equal to the velocity dispersion 
in the inner region of the system. 
Observational data show that regions close to the Sagittarius globular clusters
(Terzan 8, Terzan 7, Arp 2) 
have a similar projected velocity dispersion
(Ibata et al. \cite{Ibata97}). 

 The density contour map of the s-B2b model at 8.47 Gyr (Fig. \ref{f14}c)
resembles the observational map for the region around the center of
the satellite (see for example Fig. 1 in Ibata et al. \cite{Ibata97}),
in spite of the low
resolution (few mass points) of the simulation. However, the center of our 
satellite has a steeper density profile, the half-brightness
radius of the model is $r_{hb}=0.12$ kpc and that of Sagittarius DSph is
$0.55$ kpc. Even at the end of the simulation ($t=10.5$ Gyr)
the satellite remains too concentrated ($r_{hb}=0.17$ kpc). This
discrepancy could mean that
either the real Sagittarius DSph has undergone a
stronger tidal field, which has caused an effective destruction of the 
satellite core, or it has suffered tidal disruption for a longer time,
or the initial density profile of the Sagittarius satellite was
more extended 
than that of our model. In order to test the latter hypothesis, we 
have built the s-B2c
model, which is less concentrated than s-B2b but more massive. s-B2c follows the
same orbit as s-B2b.

{\it s-B2c model:} 

 The {\bf s-B2c satellite} survives 5.5 Gyr before disruption. The 
effects at 
perigalacticon (Fig. \ref{f12}c) are stronger 
than those of the other s-B2 models, 
because of the lower concentration. 
Low surface brightness trailing and leading streams are formed. 
At 5.88 Gyr, the residual satellite looks like a long tidal extension
with a maximum projected surface brightness $\mu \sim 27.0$ mag arcsec$^{-2}$,
assuming $(M/L)_{\mbox{real}} = 2$ M$_{\odot}/$L$_{\odot}$ (Fig. \ref{f13}c).

 The line-of-sight velocity dispersion evolves as shown in 
Fig. \ref{f11}c. 
In the outer region of the satellite
it increases up to an almost constant velocity dispersion along the orbit. At 
the last snapshot of Fig. \ref{f11}c 
(5.88 Gyr) the satellite is already disrupted 
and the velocity dispersion of the remnant stream has increased 
with respect to the value before destruction.

 The central $(M/L)_o$ ratio obtained from Eq. (\ref{ML}) depends 
on the position of the satellite along the orbit (Fig. \ref{f10}). 
The most important variations in the $(M/L)_o$ ratio are observed at
the perigalactic passage. At this time the satellite is compressed 
and stretched, its internal distribution is modified and 
the tidal extensions are more pronounced. At the final steps of 
the evolution, 
the more stripped the satellite is, the higher measured the $(M/L)_o$ is. 

\begin{figure}
 \resizebox{\hsize}{!}{\includegraphics{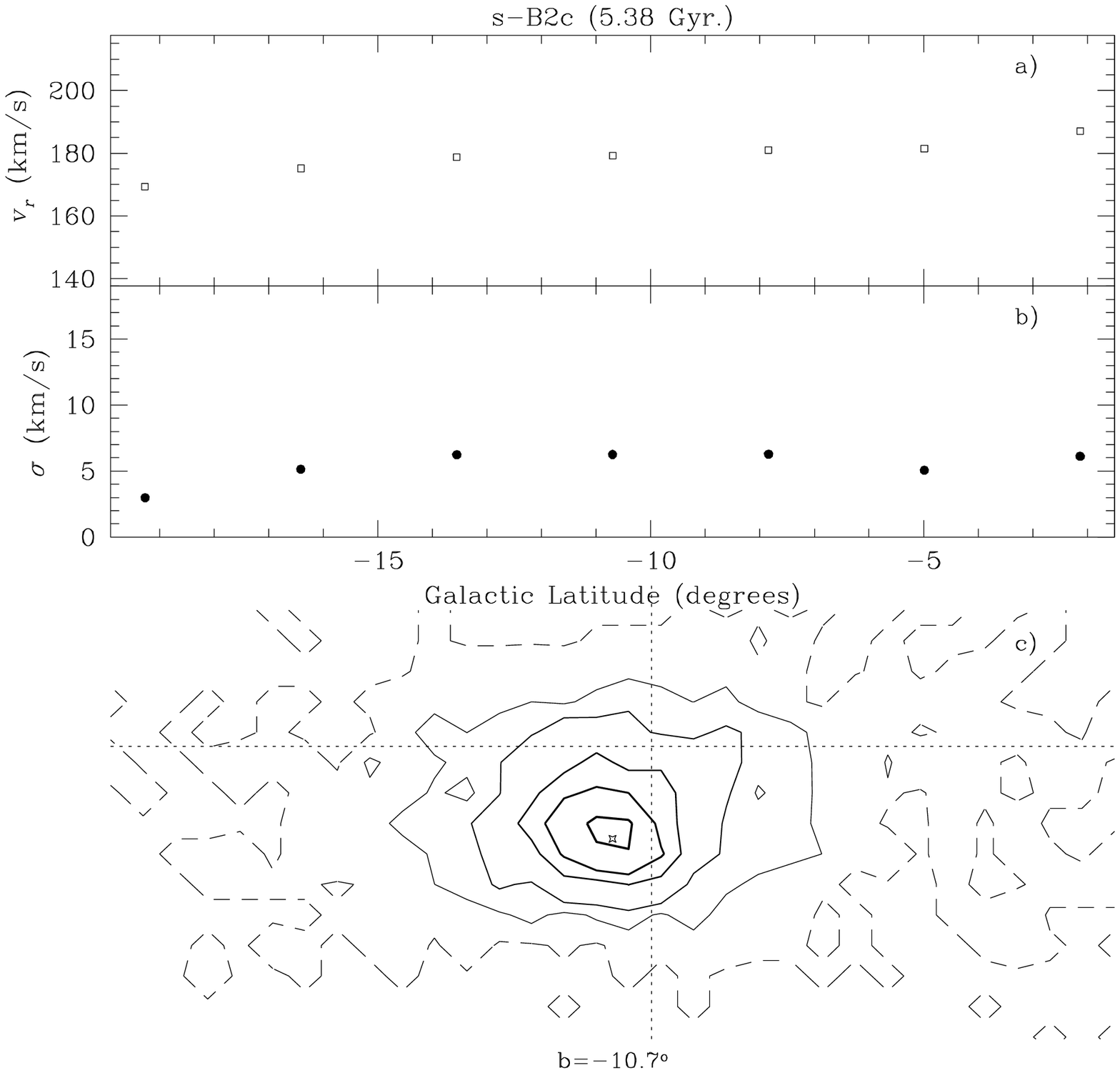}}
  \caption{s-B2c satellite, at 5.38 Gyr: 
           {\bf a} Radial velocity, $v_r$, along the orbit (constant galactic 
	   latitude),
           {\bf b} velocity dispersion, $\sigma$, along the orbit 
           and {\bf c} contour density map (the star represents the mass 
           center of the satellite, located at $b=-10.7^o$), 
           the 
           solid contours correspond to $\mu=24.3, 25.0, 25.5 26.7, 27.5$ mag arcsec$^{-2}$ 
           from the center 
           and
           the dashed contour corresponds to $\mu=30$ mag arcsec$^{-2}$}
  \label{f15}
\end{figure}

 We have analyzed the satellite characteristics at 5.38 Gyr from the beginning 
of the simulation. It corresponds to the last approach to the galactic
disc and after the last perigalacticon, which produces total disruption. At 
this time, the position ($b=-10.7^o$) 
and the Galactic velocity 
($(U,V,W)\sim(214,0:,170)$ km s$^{-1}$) 
of the
satellite agree with the observational data. In Fig. \ref{f15},
the kinematical distribution (radial velocity 
and line-of-sight velocity dispersion) along the orbit and
the contour density maps are shown. The half-brightness
radius, $r_{hb}$,  and central surface brightness, $\mu_o$, of the model 
roughly fit the observations (Table \ref{Sgr_tbl}).
The values we obtain in the \hbox{s-B2c} simulation at 5.38 Gyr 
are $r_{hb} \sim 0.4$ kpc and $\mu_o \sim 24.3$ mag arcsec$^{-2}$, which are
slightly lower than observational values, but 
they evolve to larger half-brightness radius and lower surface brightness as
the satellite approaches the Galactic disc and gets more disrupted. For 
example, at $t=5.565$ Gyr we obtain the best accord between the simulations
and the observations for $r_{hb}$ and $\mu_o$ (the values of the model are
0.5 kpc and 25.4 mag arcsec$^{-2}$, respectively), but the position and
the velocity do not reproduce the observations ($b\sim 20^o$ and $(U,V,W) \sim
(103,0,86)$ km/s).

 The radial velocity gradient along the orbit of s-B2c satellite depends on 
the orientation of the trajectory, as in the s-B2b model. At 5.38 Gyr
we obtain $|dv/db| \sim 1.5$ km s$^{-1}$/degree for the trailing
stream, which is lower than the observed value. However, the radial velocity
$v_r$ is in agreement with the observations ($v_r=171$ km s$^{-1}$). 
The satellite velocity dispersion 
measured inside 0.5 kpc 
has decreased 
($\sigma \sim 7$ km/s), but it is still large enough to give $(M/L)_o \sim 10
(M/L)_{\mbox{real}}$.

\begin{figure}
 \resizebox{\hsize}{!}{\includegraphics{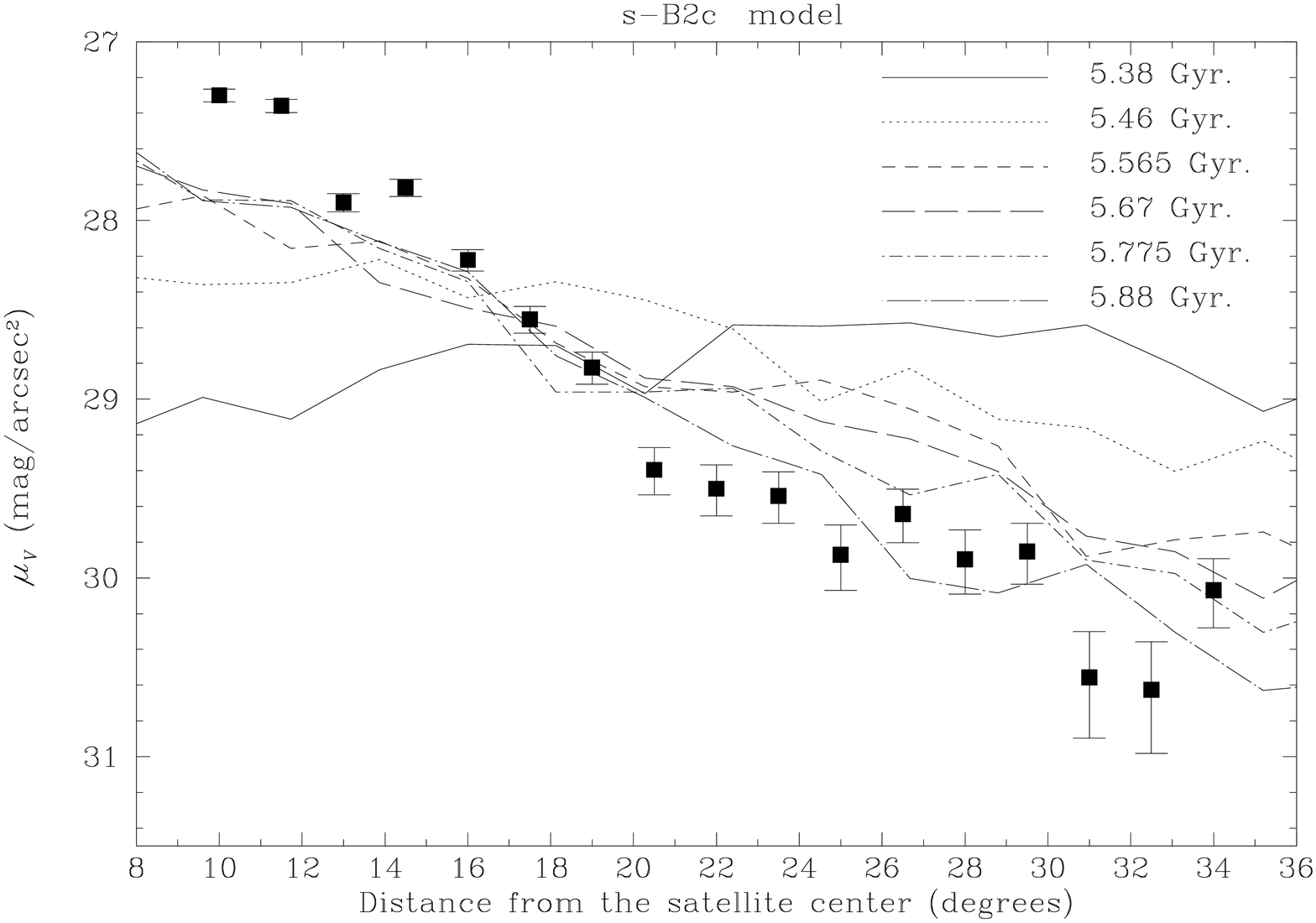}}
  \caption{Surface brightness $\mu_V$ of the tidal tail close to the 
           Sgr satellite.
           Lines represent the s-B2c model at various timesteps. Filled
           squares are the data (with the errorbars) 
           by Mateo et al. (\cite{Mateo98})}
  \label{f16}
\end{figure}

\begin{figure}
 \resizebox{\hsize}{!}{\includegraphics{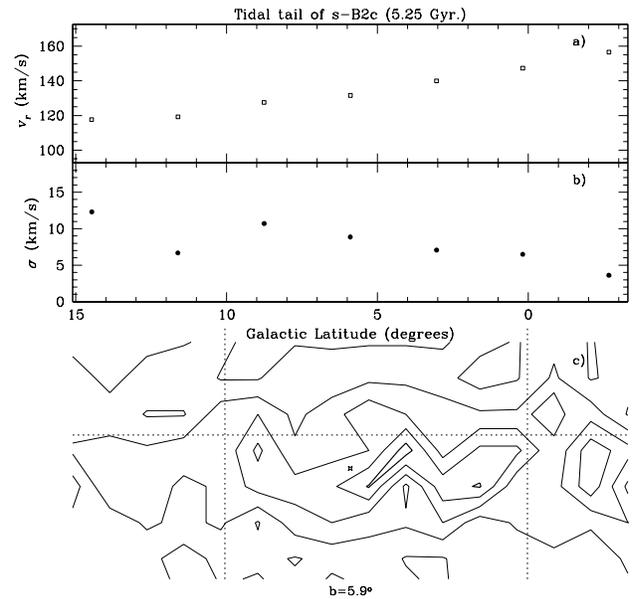}}
  \caption{Tidal tail of the s-B2c satellite: 
           {\bf a} Radial velocity, $v_r$, along the orbit (constant galactic
	   latitude),
           {\bf b} velocity dispersion, $\sigma$, along the orbit and
           {\bf c} contour density map (the star represents the mass center
           of the satellite, located at $b=5.9^o$) and the contours correspond to
           $\mu = 29.6, 29.9, 30.3, 30.8, 32.0$ mag arcsec$^2$}
  \label{f17}
\end{figure}

 It is interesting to analyze not only the region around the center of 
mass of the satellite but also the tidal tails which are formed along the
orbit. We have done it for several tidal tails close to the satellite. 
In Fig. \ref{f16}, we plot
the surface brightness of the trailing tail of the s-B2c model 
at various timesteps, assuming 
$(M/L)_{\mbox{real}} = 2$ M$_{\odot}/$L$_{\odot}$ (a larger value of $(M/L)_{\mbox{real}}$ moves 
the lines down in the figure and vice versa). The simulations are compared
with the trailing tail observed by Mateo et al. (\cite{Mateo98}) 
recently. As we can see, except for the region closest to the satellite 
center 
($<12^o$),
the simulations reproduce the observations, 
giving a better result when the satellite is more disrupted. This supports
the hypothesis that Sgr DSph is close to its destruction. In our simulations, 
we also obtain a leading tail similar to the trailing one. Unfortunately, 
the proximity of the Galactic disc to Sgr center
makes difficult to test this result of symmetric tails
due to still poor observational data. 
To give another example of the complexity of the satellite structure,
we plot in Fig. \ref{f17} the contour density levels and the 
kinematic behaviour
of a tidal tail, at 5.25 Gyrs from the beginning of the simulation.  
It looks like the Sgr DSph
from the kinematical point of view, having radial velocity (Fig. 
\ref{f17}a) and 
velocity dispersion (Fig. \ref{f17}b) similar to Sagittarius observed values
($v_r = 171$ km s$^{-1}$ and $\sigma = 11.4$ km s$^{-1}$).
The shape of the tail (Fig. \ref{f17}c) also looks 
like a tidally disrupted satellite, however, 
the central surface brightness is $\mu_o \sim 29.5$ mag arcsec$^{-2}$ (fainter
than a typical DSph) and the computed $(M/L)_o$ ratio is $200 (M/L)_{\mbox{real}}$ 
(higher than the estimation for 
DSph galaxies). Nevertheless, this example illustrates,
on one hand, the difficulty of distinguishing between a tidal tails and a
DSph galaxy when the latter is close to the disruption and, on 
the other hand, the 
possibility of identifying objets (globular clusters, etc) which have
been tidally disrupted from a satellite galaxy by measuring their kinematic
characteristics.

\section{Discussion}

 Summing up the results obtained in our simulations and comparing them 
with other published works, we agree with other authors (Vel\'azquez
\& White \cite{Velazquez95}; Ibata \& Lewis \cite{Ibata98}) 
that Sgr DSph has a short period orbit
($T \leq 1$ Gyr). This constraint comes from measurement of the 
position and the radial velocity of Sgr DSph in a realistic model of the
Galaxy.
However, we disagree about the DM content. These
previous works claim that Sgr DSph is a DM dominated satellite,
because, contrary to us, they could not obtain a low mass 
satellite which survives by orbiting in the potential of the Galaxy.
Therefore, they 
cannot explain the complex evolutionary pattern inferred for the
Sagittarius DSph, which has suffered a chemical 
enrichment and evolution and the age interval of its globular
clusters which suggests a long life orbiting in the Galaxy.

Ibata \& Lewis (\cite{Ibata98}) have tested low massive models in order 
to reproduce
the Sgr DSph characteristics, reaching an unsuccessful result on the 
matter and concluding that Sgr DSph must have a large DM content. 
There are two differences between their 
study and ours. Firstly, they use a rigid potential to model the Galaxy, so
no energy interchange is allowed between the satellite and 
the Galaxy. This restriction could eventually 
prevent a readjusting of the 
internal energy
of the satellite in order to reach an equilibrium with the environment.
Even if the energy transfer from the satellite to the halo is not 
important in all cases, we emphasize that 
our treatment is {\it a priori} adaptable to 
satellites of various masses.
Secondly (and mainly), 
Ibata \& Lewis (\cite{Ibata98}) do not take into account the tidal
potential of the Galaxy as they build their satellite model.
We have confirmed that 
the fate of such satellites is different from that of the satellites in 
equilibrium with the tidal potential as considered here. 
Consequently, a more accurate model
of the initial satellite, that reflects the true dynamical situation
of the DSph, is required in order to avoid spurious effects in the simulations.
These differences in the models are responsible for the opposite conclusions
reached by Ibata \& Lewis, compared with ours concerning
DM. Furthermore, the recent observations of Sgr by Mateo et al. 
(\cite{Mateo98}) prove the existence of a long trailing tail, which supports the
hypothesis that Sgr DSph has suffered strong tidal forces, it is close
to disruption and, therefore, it is not in virial equilibrium. The existence 
of this tail also reduces the inferred $(M/L)_o$ ratio.

Our low DM satellites ({\it modified King's models}) 
establish important restrictions on the satellite formation theory. Thus, 
the satellite galaxies grow either in a quasi-isolated region
(falling slowly on the center of the primary galaxy and having time for 
readjusting their internal structure to the tidal forces), or inside 
the main galaxy, in the tidal tails of a major accretion event, which
automatically implies equilibrium with the tidal forces at the satellite 
formation epoch. Furthermore, the survival time of the {\it equilibrium} 
satellites in the 
potential of the Galaxy depend on the initial concentration of the DSph (the
larger concentration, the longer the life-time). We have obtain models (i.e.,
s-B2a) which survive more than 10 Gyr. The evolution of these satellites
gives rise to tidal streams and modifications on the outer velocity 
dispersion of the satellite and it leads to high observed
mass-to-light ratios if dynamical equilibrium is assumed. 
We have obtained a rather good qualitative 
agreement with the observational
constraints for the model s-B2c in a time interval (from $5.38$ to $5.7$ Gyr),
although this agreement is not achieved  simultaneously on all the
constraints.
We remind that the simulation time does not necessarily correspond with
the age of the oldest globular cluster of Sgr and, therefore, the s-B2c
model is not necessarily in disagreement with the observations. 
In this case, two
possibilities could be invoked to explain the age of the oldest Sgr globular
clusters: (i) either the satellite
has been orbiting for a long time in a more external region (where tidal forces are smaller), it suffers dynamical friction or/and a deflection with 
a dense structure of the MW and it reaches the present orbit where, eventually, 
it is disrupted, or (ii) it has formed in a tidal
tail which contained an old stellar population (Kroupa \cite{Kroupa98c}).

The tidal tail models illustrate that a DSph without DM orbiting through 
central regions of the Galactic potential could 
preserve its structural parameters for a long time. 
Later on, the DSph could show high inferred $(M/L)_o$ values as it becomes 
disrupted. The only restriction of the model is that the satellite must
reach the central regions of the Galaxy in equilibrium with the tidal 
forces. That could be possible whether the satellite has been formed in a 
tidal tail of an accretion event or it has been slowly accreted
from the external parts of the Galaxy. The latter possibility could increased 
the estimated life-time of the satellite before destruction. 

Recently, K98 has discussed the parameter space of the
DSphs which could survive in a tidal field. In spite of the 
different initial distribution functions, our s-models 
seem to be
located in the available region of the ($M_{sat}$, $r_{sat}$) space for 
surviving satellites. However, the main difference between
the approaches to the problem is that we build the model of the 
satellite initially in equilibrium with the tidal forces by solving the 
Boltzmann equation (G\'omez-Flechoso \& Dom\'{\i}nguez-Tenreiro 
\cite{Gomez97}) and that assures {\it a priori} the survival of the satellite
for a long time. 

\section{Summary}

 We have performed simulations of Sagittarius-like satellites to follow
the evolution of these systems. 
The satellite and the primary galaxy are modelled as N-body
systems, that prevents possible purely numerical effects which could appear 
in a non-selfgraviting scheme (i.e. if one of the interacting 
system is represented by a rigid potential). Two satellite models 
have been tested:
\begin{enumerate}

 \item First of all, we have used {\it isolated King's models} (three 
free parameter models) for the satellite, but they are tidally stripped and
destroyed in a short interval of time, undergoing a fast evolution. 

 \item In order to correctly model 
the observed situation we had to introduce the 
{\it modified King's models}
(two free parameters-models), 
which allow us to build up the distribution
function of the satellite, taking into account the tidal force of the 
environment. These s-models could correspond to no DM dominated satellites. 
 In this framework, 
the initial concentration of the satellite 
and the trajectory determine its life-time. Central circular orbits lead to
more effective tidal destruction events than highly eccentric orbits. 
For a given 
kind of orbit, the most concentrated satellites live longer. 
The model
which better reproduces the observations is the s-B2c model after
orbiting 5.38-5.6 Gyr in the potential of the Milky Way. 
For the same orbits, the s-models survive a longer time orbiting
in the Galactic potential than the f-models, despite the
s-models having smaller masses (smaller binding energies). 

\end{enumerate}

 According to the results presented here, 
our 
main conclusions about the evolution of Sgr are:
\begin{enumerate}

\item Sagittarius 
may not be 
a dominated DM satellite.

\item It follows an eccentric orbit (perigalacticon and apogalacticon
are approximately 15 and 70 kpc).

\item It was more concentrated in the past than
at the present epoch. 

\item It could have 
been orbiting in the Galactic halo for a long 
time (minimum 5 Gyr).

\end{enumerate}

As general conclusions about the DSph satellites, we can remark that:

\begin{enumerate}
\item High values of $(M/L)_o$ found in the literature (Ibata et al. \cite{Ibata95};
Irwin \& Hatzidimitriou \cite{Irwin95}; Mateo \cite{Mateo98b} 
for a review of the Galactic 
satellites) could be due to an erroneous use of the
virial theorem and not to the presence of DM (as 
already suggested by K97;
KK98). 

\item The progenitor of these satellite could be either lumps formed in a 
major accretion event or other dwarf galaxies (maybe dwarf irregulars or
dwarf spirals) which fall in the 
potential well of the Galaxy, lose their gas content and evolve to the 
equilibrium configuration in the tidal field of the Milky Way, giving rise
to the population of DSphs. The finding of intermediate states of this 
evolution could be a support to the proposed scenario.

\item According to some of the present results, some DSphs are able to
survive for long times orbiting in the Galactic halo.
That should be taken into account for the calculations of the dissolution 
rate of dwarf galaxies
in the halo of the primary galaxy. 

\end{enumerate}

\begin{acknowledgements}

We would like to thank D. Pfenniger and D. Friedli for critically reading
the paper, as well as the referee for remarks which have led to
clarify various points of the paper. 
M.A. G\'omez-Flechoso was supported by the Fundaci\'on Ram\'on Areces 
through a fellowship. This work has been partially supported by the Swiss
National Science Foundation. 

\end{acknowledgements}

\end{document}